\documentclass[conference]{IEEEtran}
\usepackage{epsfig,graphicx,subfigure}

\usepackage[table]{xcolor}
\usepackage{graphicx}
\usepackage{amsmath}
\usepackage{epsfig}
\usepackage{xspace}
\usepackage{multirow}
\usepackage{amsfonts}
\usepackage{tikz} 

\usepackage{graphicx,subfigure}
\usepackage{wrapfig}
\usepackage{color}
\usepackage{url}
\usepackage{rotating}
\usepackage{tabularx}

\usepackage{soul,color}
\usepackage{cite}
\usepackage{rotating}
\usepackage{pdflscape}
\usepackage{tablefootnote}
\usepackage[noend]{algpseudocode}
\usepackage{algorithm}
\usepackage{hyperref}

\newtheorem{definition}{Definition}
\ifCLASSINFOpdf
\else
\fi
\begin{document}
%
\title{Detecting Overlapping Communities from Local Spectral Subspaces }

\author{
\IEEEauthorblockN{Kun He  \quad ~~  Yiwei Sun}
\IEEEauthorblockA{School of Science and Technology\\
Huazhong~University of Science and Technology\\
Wuhan 430074, China\\
Email: \{brooklet60, yiweisun\}@hust.edu.cn}
\and
\IEEEauthorblockN{David~Bindel~~~~~~John~Hopcroft~~~~~~Yixuan Li}
\IEEEauthorblockA{Department of Computer Science,\\Cornell University\\
Ithaca 14850, NY, USA\\
Email: \{bindel, jeh, yli\}@cs.cornell.edu}
}


%


\maketitle

\begin{abstract}
Based on the definition of local spectral subspace, we propose a novel approach called \emph{LOSP} for local overlapping community detection.
Instead of using the invariant subspace spanned by the dominant
eigenvectors of the entire network, we run the power method for a few
steps to approximate the leading eigenvectors
that depict the embedding of the local neighborhood structure around seeds of interest.
We then seek a sparse approximate indicator vector in the local spectral subspace spanned by these vectors
such that the seeds are in its support.

We evaluate \emph{LOSP} on five large real world networks across various domains with labeled ground-truth communities and compare the results with the state-of-the-art community detection approaches. \emph{LOSP} identifies the members of a target community with high accuracy from very few seed members, and outperforms the local Heat Kernel or PageRank diffusions as well as the global baselines.

Two candidate definitions of the local spectral subspace are analyzed,
and different community scoring functions for determining the community boundary, including two new metrics, are thoroughly evaluated.
The structural properties of different seed sets and the impact of the seed set size are discussed. We observe
low degree seeds behave better, and \emph{LOSP} is robust even when
started from a single random seed.

Using \emph{LOSP} as a subroutine and starting from each ego connected component,
we try the harder yet significant task of identifying all
communities a single vertex is in. Experiments show that the proposed method achieves high F1
measures on the detected multiple local overlapping communities containing the seed vertex.~\footnote{A shorter version is accepted by ICDM2015.}

\end{abstract}


%
\IEEEpeerreviewmaketitle

\section{Introduction}
\label{sec:Introduction}

With the rapid growth in the size of networks,  exploring the global community structure~\cite{Fortunato2012survey,Papadopoulos2012survey,coscia2012demon,Ahn2010LinkCommunities} 
becomes more costly and inefficient.
A growing body of interest has shifted the attention from the macroscopic structure to the microscopic structure in mining the local community structure of interest
 ~\cite{IsabelKDD14,kloster2014heat,FreeRider2015,JaewonICDM2012}.

In many contexts, with the information of a few exemplary community members
shared by some ``domain experts", we are interested in finding the
latent members in the target community. The context of finding local community structure can be widely applied to many real world scenarios. For example, in political participation networks, one might discover
the political group membership from a small set of representative politicians
\cite{polarization2013}. 
Also, in web search, from a few web pages that share similar information,
we could generate a larger group of web pages that contain the relevant information with respect to a certain search query. In product networks, sales websites may recommend potential products to the customer by expanding from a few purchased products based on the co-purchase networks, and generate a larger set of
products that might be of interest to the customer. Furthermore, in biological
networks representing the interactions among genes, biologists are likely to discover a set of genes that form a functionally similar unit  starting from  a few observed and well-studied genes.  

Some existing approaches based on global seed set expansion grow each seed set into a larger set of vertices by locally optimizing a
community scoring function
such as conductance or modularity~\cite{Pan2012seed,JaewonICDM2012,kloster2014heat,FreeRider2015}. These algorithms do not scale very well in terms of running time and memory consumption since they heavily rely on the whole graph structure.
 Recent work has adapted these global seed set expansion approaches to the local community detection problem.
 This type of local expansion algorithm includes, for example, the personalized PageRank diffusion~\cite{JaewonICDM2012,IsabelKDD14}
 and the Heat Kernel diffusion~\cite{Chung2007heat,Chung2013heat,kloster2014heat}, both of which find local communities by utilizing the dynamics of random walks starting from the seeds.

The random walk technique has been extensively adopted
as a subroutine for locally growing the seed set in the literature. Most existing literature on community detection has been focusing on the single probability vector after short random walks~\cite{JaewonICDM2012,IsabelKDD14,Chung2007heat,Chung2013heat,kloster2014heat}.
Based on our previous investigation on local spectral clustering~\cite{li2015uncovering},
we provide a systematic approach for finding small, overlapping communities using the subspace shaped by the dynamics of short random walks, which we call  \emph{LOSP (Local Spectral)}.

 The first part of this work will address fundamental questions concerning the local
structure in large networks such as, how do we find a local community in
time that is a function of the size
of the target community rather than the size of the entire network?
 How can we effectively find local structure without getting access to the entire network?
 Do we choose a good stopping rule when growing the seed set to a local community?
 How do the quality and quantity of the seed members affect the performance of \emph{LOSP}?
We also consider \emph{LOSP} when the ``domain expert" poorly selects the initial seed members.

In the second part of this paper, we tackle the problem of identifying all the communities a single vertex is in, which we refer to as the \emph{multiple membership identification} problem. Identifying all local communities is useful in numerous applications. For instance, social network users may be interested in exploring all social groups an individual is in, and biologists would like to mine all GO (Gene Ontology) terms a gene serves.

We explain the classical spectral clustering method before presenting \emph{LOSP}.
Spectral clustering starts with computing the first
$d$ eigenvectors of the graph Laplacian or some other matrix associated with the network.
Each row in the eigenvector matrix corresponds to the embedding of a vertex in a $d$-dimensional space. These vectors are then clustered using some method such as $k$-means, resulting in disjoint communities. Each community could be further partitioned until communities of the desired size are found.
This method is not likely to work well if the communities are small and
many partitions are needed. Moreover, it cannot handle the task of overlapping
community detection.

We make two fundamental changes to the  current spectral
clustering method:

\begin{itemize}
\item \textbf{Handle the overlapping situation}.
Instead of clustering the row vectors in the $d$-dimensional subspace, we
seek a minimum $\ell_{1}$ norm indicator vector in the space spanned
by the eigenvectors containing the initial seeds.
Overlapping communities correspond to different seed sets.
\item \textbf{Define the local spectral subspace}.
We calculate the first few eigenvectors by the power method, but
instead of iterating until it converges, we iterate for a few steps
such that the probability distribution of a random walk starting from the seeds would reach out to the vertices of the local community but not spread out to the entire graph.
\end{itemize}

To have a comprehensive understanding of the network structure and
the proposed method, we thoroughly analyze variations of \emph{LOSP}
and explain the intuition behind the method. Five community scoring
functions are evaluated, including new definitions of triad-participation-number (TPN) and normalized modularity.
\emph{LOSP} is robust in showing little fluctuations on different scoring functions, but TPN and conductance yield the best results. We also
investigate the structural properties of different seed sets, and
find that low degree seeds and random seeds are essentially the same
for real world networks in finding the local structure.

The performance evaluation on various real-world networks shows that \emph{LOSP} outperforms the prevalent PageRank or Heat
Kernel diffusion methods as well as the state-of-the-art global methods.
We present a comprehensive understanding of the local spectral clustering for effectively and efficiently finding local community structure, and we believe the insight gained through this research will shed light on the area of local community detection.

\section{Related Work}
\label{sec:relatedWork}

A considerable amount of literature has been published
on finding communities in large social and information networks. We highlight a few recently emerged ideas for finding communities using seed set expansion
that are most related to our work.

\textbf{Global seed set expansion.}
Many global community detection algorithms fall into the seed set expansion category.
OSLOM, for example, starts with each vertex as the initial seed and optimizes a fitness function,
defined as the probability of finding the cluster in a random null model,
 to join together small clusters into statistically significant larger clusters~\cite{lancichinetti2011OSLOM}.
Greedy Clique Expansion (GCE)~\cite{lee2010GCE} expands maximal cliques as the seeds to optimize a local fitness function similarly to the conductance.
Seed Set Expansion (SSE)~\cite{Whang2013SSE} identifies overlapping communities
by expanding different types of seeds using a random walk with restart scheme called the personalized PageRank.

\textbf{Local seed set expansion.}
The random walk technique has been extensively adopted as a subroutine for locally expanding the seed set~\cite{andersen2006communities,Liu2014randomwalks,jin2011markov}
 and it is observed to produce communities correlated highly to the ground-truth communities in real-world networks~\cite{Bruno2014separability}.
PageRank~\cite{PageRank2006, JaewonICDM2012, IsabelKDD14} and Heat Kernel~\cite{kloster2014heat,Chung2007heat,Chung2013heat} are two main techniques for the probability diffusion.

Spielman and Teng\cite{PageRank2004} use degree-normalized, personalized PageRank (DN PageRank) with respect to the start seed and do truncation on small values,
 leading to the PageRank Nibble method\cite{PageRank2006}.
And the DN PageRank is adopted by several PageRank-based clustering algorithms~\cite{andersen2006communities,JaewonICDM2012},
 which are competitive with a sophisticated and popular algorithm METIS~\cite{METIS1995}.
Kloumann and Kleinberg\cite{IsabelKDD14} evaluate different variations of PageRank, and find that the standard PageRank yields better performance than the DN PageRank.

The Heat Kernel provides another local graph diffusion~\cite{Chung2007heat,Chung2013heat,kloster2014heat},
and involves the Taylor series expansion of the matrix exponential of the random walk transition matrix.
Chung et al. analyze the property of this diffusion theoretically~\cite{Chung2007heat}, and propose a randomized Monte Carlo method to estimate the diffusion~\cite{Chung2013heat}.
Kloster et al. propose a deterministic method that uses coordinate relaxation on an implicit linear system that estimates the Heat Kernel diffusion, and show that Heat Kernel outperforms the personalized PageRank by finding smaller sets with substantially higher F1 measures~\cite{kloster2014heat}.

There are also other local methods based on the random walk technique.
For instance, Wu et. al~\cite{FreeRider2015}
use a variant of the degree normalized, penalized hitting probability to weight the nodes by starting from the query nodes,
and define the reciprocal of the weight as the query biased density to effectively reduce the free rider effect that tends to include irrelevant subgraphs in the detected local community.

\textbf{Stopping rules for community boundary.}
All seed set expansion methods need a stopping criterion for defining the community boundary unless
the size of the target community is known.
Conductance is commonly recognized as the best stopping criterion due to its intrinsic local property
\cite{kloster2014heat,Whang2013SSE,JaewonICDM2012}.
Yang and Leskovec provide widely-used real world datasets with labeled ground truth\cite{JaewonICDM2012},
and find that conductance and triad-partition-ratio (TPR) are the two stopping rules yielding the highest detection accuracy.
The Heat Kernel method also adopts conductance as the stopping rule for the local community~\cite{kloster2014heat}.

\textbf{Seeding strategies.} The seeding strategy is a key component for seed set expansion algorithms.
GCE selects maximal cliques as the seeds~\cite{lee2010GCE}.
Whang et al. discover that an independent set of high-degree vertices, which they called ``spread hubs"
 outperforms Graclus centers, local egonets, and random seeding strategies ~\cite{Whang2013SSE}.
Kloumann and Kleinberg\cite{IsabelKDD14} compare random seeds with high degree seeds,
and discover that random seeds are superior to high degree seeds,
 and they suggest domain experts provide seeds with a diverse degree distribution.

\section{Preliminaries}
\label{sec:preliminaries}
\subsection{Problem Statement}

Let $G =(V,E)$ be a connected, undirected graph with $n$ vertices and $m$ edges,
and let $\mathcal{C}$ be a set of labeled ground truth communities.
We are interested in answering the following questions: 

\begin{itemize}
\item Given a few exemplary members $S$ in a target community $C_k \in \mathcal{C}$ where $|C_k| \ll n$,
   how can we identify the remaining latent members in $C_k$? 
\item For a single vertex $s$, what is the set of all communities that $s$ belongs to?
\end{itemize}


\subsection{Community Scoring Functions}
\label{sec:ScoringFunctions}
We provide several community scoring functions to characterize how community-like a subset of vertices is mathematically.
For a community $C_k$, let $n_k$ and $e_{kk}$ be the number of vertices and edges within the community, 
and $d_k$ be the total degree of the internal vertices.
We adopt three often-used scoring functions, modularity (Mod), conductance (Cond) and triad-participation-ratio (TPR), and define two new scoring functions, normalized modularity (nMod) and triad-participation-number (TPN) to evaluate the quality of a community.

In the experiments in Section \ref{sec:experiments_LOSP} using \emph{LOSP} to
find communities, TPN and conductance produce the communities most resembling the ground truth communities.

\subsubsection{Modularity}

Modularity 
is defined as the fraction of the internal edges of the communities minus the expected fraction if edges were distributed at random
 while preserving the degree distribution~\cite{newman2006modularity}.
The modularity $Q$ of a partitioning is calculated by:
\setlength{\belowdisplayskip}{0pt} \setlength{\belowdisplayshortskip}{0pt}
\setlength{\abovedisplayskip}{0pt} \setlength{\abovedisplayshortskip}{0pt}
\begin{equation*}\label{EqMod}
Q = \sum_{k=1}^c Q_k = \sum_{k=1}^c\left[\frac{e_{kk}}{m}-\left( \frac{d_k}{2m} \right)^2\right],
\end{equation*}
where $c$ is the number of communities.
$Q$ lies in the range $[-0.5,1)$, and is positive if the number of internal edges exceeds the number expected on the basis of chance. $Q_k$ is the modularity of community $C_k$. The larger the better.

\subsubsection{Normalized Modularity}

Instead of using ``minus" to define the modularity, we use ``divide" and define
 the normalized modularity (or normalized density) $D$ as the fraction of the edges within the communities divided by the expected fraction if edges were distributed at random. So
 $D = 4m\sum_{k=1}^c e_{kk}/d_{k}^2$.
As the coefficient $4m$ is a constant when evaluating community qualities in the same network,
we simply define
\setlength{\belowdisplayskip}{0pt} \setlength{\belowdisplayshortskip}{0pt}
\setlength{\abovedisplayskip}{0pt} \setlength{\abovedisplayshortskip}{0pt}
\begin{equation*}\label{Eq22}
D = \sum_{k=1}^c D_k = \sum_{k=1}^c \frac{e_{kk}}{d_{k}^2},
\end{equation*}
which has the advantage of being insensitive to the network scale.
$D_k$ is the normalized modularity of community $C_k$, and we define
$D_k = 0$ when both the numerator and the denominator are 0 indicating only one isolated vertex in this community.
$D_k > \frac{1}{4m}$ if the fraction of internal edges exceeds the expected fraction.

\subsubsection{Conductance}



\emph{Conductance} is a concept from physics that controls how fast a random walk on a graph converges to the stationary distribution\cite{John2015}.
As we are looking for small communities in large networks, we assume the size of each community is much less than half of the network.
Thus, the conductance of a community $C_k$
\setlength{\belowdisplayskip}{0pt} \setlength{\belowdisplayshortskip}{0pt}
\setlength{\abovedisplayskip}{0pt} \setlength{\abovedisplayshortskip}{0pt}
\begin{equation*}\label{EqCond}
\Phi_k =\frac{d_k - 2e_{kk}}{d_k} = 1- 2\frac{e_{kk}}{d_k}
\end{equation*}
is the fraction of total edge volume leaving the community\cite{conductance2000}.
The lower the better.

%
%
%

\subsubsection{TPR and TPN}

Triad-participation-ratio (TPR) is defined as the fraction of vertices in the community that belong to triads\cite{JaewonICDM2012}.

Based on TPR, we define another scoring function triad-participation-number (TPN) as
the average number of triangles a vertex belongs to.
For each vertex $v$ in community $C_k$,
  consider its one-step neighborhood, namely its ego network excluding the ego. The number of edges in $egonet{_{\neg{v}}}(v)$ is just the number of triangles in $egonet(v)$.
  Divide the value by 3 as each triangle is calculated three times.
\setlength{\belowdisplayskip}{0pt} \setlength{\belowdisplayshortskip}{0pt}
\setlength{\abovedisplayskip}{0pt} \setlength{\abovedisplayshortskip}{0pt}
$$ TPN_k = \frac{ \sum_{v\in C_k} Vol(egonet{_{\neg{v}}}(v))}{3n_k} $$

TPR and TPN are based on the internal connectivity, while modularity, normalized modularity and conductance are based on the internal and external connectivity.
By a community scoring function, one sets a threshold to define a local community using a community scoring function
\footnote{One could also use modularity or other functions mentioned above (the larger the better), and define $C$ as a local community if its scoring value is no less than a parameter $\delta > 0$.}.

\begin{definition}
\textbf{Local Community}. A subset $C \subset V$ with $|C| \ll |V|$ is called a local community evaluated by conductance with parameter $\epsilon>0$ if it satisfies  $\Phi(S) \leq \epsilon$.
\end{definition}

\vspace{-0.5em}
\subsection{Datasets}

For the performance evaluation,
we consider five real-world network datasets from the Stanford Network Analysis Project (SNAP)\footnote{\url{http://snap.stanford.edu}}.
For each network, we adopt the top 5000 annotated communities with the highest quality evaluated on several metrics by Yang \& Leskovec\cite{JaewonICDM2012} as the ground truth communities.

Table \ref{real_data_summary} summarizes these datasets and the statistics on the size and conductance of the ground truth communities.
Values for diameter $\mathcal{D}$ (length of the longest pairwise shortest paths) and 90-percentile effective diameter $\mathcal{D}_{90\%}$ are from the SNAP website.
Each of these datasets approximately follows a power law degree distribution $P(d) \sim d^{-\mu}$.

For Amazon, vertices represent products, vertices of co-purchased products are connected by an edge, and
the ground truth corresponds to product categories.
For DBLP, vertices represent authors, edges represent the co-authorship,
 and the ground truth communities correspond to conferences.
The other three social networks indicate the friendships among users and the ground truth communities are user-defined groups.

\begin{table*}[htbp]
\begin{center}
\scalebox{0.9}{
\begin{tabular}{l | l  r r  c c r r| r  c }
\hline
            & \multicolumn{7}{c}{\bf{Network}} \vline& \multicolumn{2}{c}{\bf{Ground truth communities}} \\
\bf{Domain} &\bf{Name} &\bf{\# Vertices $n$}&\bf{\# Edges $m$} & $\mathbf{\mu}$ & \bf{$log(n)$ }& \bf{$\mathcal{D}_{90\%}$} & \bf{$\mathcal{D}$} &\multicolumn{1}{c}{\bf{Avg. $\pm$ Std. Size }} & \bf{Avg. Cond.}\\
 \hline
\bf{Product} &Amazon     & 334,863    & 925,872    &   2.32  &    5.52  &  15.0  & 44  &13.49 $\pm$ 17.51 & 0.07\\
\bf{Collaboration}& DBLP & 317,080    &1,049,866   &   1.03  &    5.50  &  8.0   & 21  &22.44 $\pm$  201.08& 0.41 \\
\bf{Social} &LiveJ       & 3,997,962  & 34,681,189 &   0.14  &    6.60  &  6.5   & 17  &27.80 $\pm$  58.04 & 0.39\\
\bf{Social} &YouTube     & 1,134,890  & 2,987,624  &   1.80  &    6.05  &  6.5   & 20  &14.59 $\pm$  60.46 & 0.80\\
\bf{Social}& Orkut       & 3,072,441  & 117,185,083&   0.67  &    6.49  &  4.8   & 9   &215.72 $\pm$  320.55& 0.73\\
\hline
\end{tabular}}
\end{center} 
\caption{\bf Statistics for real-world networks and their ground truth communities.}
\label{real_data_summary}
\end{table*}


\subsection{Evaluation via Ground Truth}

We adopt precision, recall and F1 score to measure how close the community $C$ expanded from a seed set $S$ is to the target ground truth community $T$ containing $S$.

The precision and recall are defined as:
\begin{equation*}
P(C,T) = \frac{|C\cap T|}{|C|}, R(C,T) = \frac{|C\cap T|}{|T|}.
\end{equation*}
And F1 score is the harmonic mean of precision and recall:
\begin{equation*}
F_1(C,T) = \frac{2 \cdot P(C,T) \cdot R(C,T) }{P(C,T) + R(C,T) } = \frac{2 |C \cap T|}{|C| + |T|}.
\end{equation*}

\section{The Local Spectral Subspace}
\label{sec:LOSPmethod}
We define a local spectral subspace (\emph{LOSP}) in order to identify the remaining latent members from very few exemplary seed members in large networks. 
In this section, we show how the \emph{LOSP} method evolves from
the classical spectral clustering method.

Let $\mathbf{A}$ be the adjacency matrix of a network $G$, $d$ the vector of vertex degrees, and $\mathbf{D} = diag(d)$ the diagonal matrix of degrees.
Add a self loop on each vertex of $G$ to get a modified graph $G'$. Let $\mathbf{ \bar A} = \mathbf{A}+\mathbf{I}$ and $\mathbf{\bar D}$ be the diagonal matrix of degrees for $G'$.
For simplicity of notation, we
use $ \mathbf{A}$ and $ \mathbf{D}$ to represent $\mathbf{ \bar A}$ and $\mathbf{ \bar D}$ in the following.

\subsection{Finding Global Disjoint Communities}
\label{sec:3_1}

 Let $\mathbf{N_{rw}}=\mathbf{D}^{-1}\mathbf{A}$ denote the transition matrix,
 and $\mathbf{N_{sym}}$ $=\mathbf{D^{-\frac{1}{2}}AD^{-\frac{1}{2}}}$ the normalized adjacency matrix.
 Let $\mathbf{V_d} \in \mathbb{R}^{n \times d} $ be a matrix containing the first $d$ eigenvectors 
 of $\mathbf{N_{rw}}$ or $\mathbf{N_{sym}}$ as the columns.
Cluster the points corresponding to the rows of $\mathbf{V_d}$
 and get disjoint communities where community
 $$C_k = \{\text{vertex}~j| \text{point}~j \in k^{th}~\text{cluster}\}.$$

\begin{figure}[ht]
\vspace{-1em}
\begin{center} \includegraphics[width=1.8in]{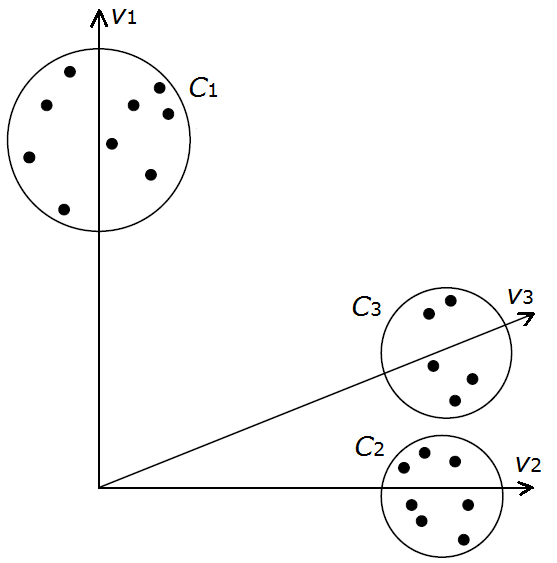} \end{center}
\vspace{-2em}
\caption{Clustering on the row vectors of $\mathbf{A}$.}
\label{fig:spectralclustering}
\end{figure}

Consider a small network with three communities that are internally cohesive with sparse mutual connections.
Form a matrix of size $n$ by 3, which consists of
an orthonormal basis for the subspace spanned by the first three eigenvectors.
Applying $k$-means clustering algorithm on the
invariant subspace basis
would result in three communities as shown in Fig. \ref{fig:spectralclustering}.

However, in most real networks, a vertex usually belongs to more than
one community.
Spectral clustering can
only partition vertices
into disjoint communities
and is unable
to handle the overlapping
detection task.  To resolve this issue, we look for the
vector $\bf{y}$ with minimum $\ell_1$ norm
in the space spanned by the leading eigenvectors to approximate
invariant subspace where the seeds are in its support.

\subsection{Defining the Local Spectral Subspace}
\label{sec:localSpectralSpace}

Consider a short random walk starting from the known seed members,
 and approximate the first $d$
 eigenvectors
 to characterize the embedding of the local network structure surrounding the seeds.

Let $\mathbf{p^{(t)}}$ be a column vector with a component for each vertex specifying the probability mass of the vertex at step $t$,
and calculate
a basis for a local approximately-invariant subspace.
\setlength{\belowdisplayskip}{4pt} \setlength{\belowdisplayshortskip}{4pt}
\setlength{\abovedisplayskip}{4pt} \setlength{\abovedisplayshortskip}{4pt}
\begin{enumerate}
\item  The initial probability $\mathbf{p^{(1)}}$  is assigned by evenly distributing the total probability among the seed members.
\item  Conduct $d-1$ steps
of the random walk
$\mathbf{N_{rw}^Tp^{(t)}} = \mathbf{p^{(t+1)}}$ to get the span of $d$
successive probability vectors\footnote{The random walk is
somewhat lazy as we added a self loop to each vertex.}, and find their orthonormal basis
$\mathbf{V_{d}^{(0)}}$, the
initial invariant subspace approximation.
 $$\mathbf{V_{d}^{(0)}} = orth([\mathbf{p^{(1)}},\mathbf{p^{(2)}},...,\mathbf{p^{(d)}}])$$
\item  Conduct $k$ steps of the random walk,
and find the orthonormal basis $\mathbf{V_{d}^{(k)}}$ using the following recurrence for the subspace iteration.
$$\mathbf{V_{d}^{(k)}} = orth(\mathbf{N_{rw}^T}\mathbf{V_{d}^{(k-1)}})$$
\end{enumerate}

The subspace dimension $d$ and the short random walk steps $k$ are some modest parameters empirically determined in Section \ref{sec:initialCommunity}.
The orthonormal basis $\mathbf{V_{d}^{(k)}}$ is what we call
the \emph{local spectral basis}
and the subspace spanned by the basis is
called the \emph{local spectral subspace}.

\subsection{Finding Local Overlapping Communities}

With the local spectral basis
$\mathbf{V_{d}^{(k)}}$, 
we look for row vectors in the spanned subspace that are nearly in the same direction as the seeds.
Mathematically, it is equivalent to solve the following linear programming problem $LP_1$:
 \setlength{\belowdisplayskip}{4pt} \setlength{\belowdisplayshortskip}{4pt}
\setlength{\abovedisplayskip}{4pt} \setlength{\abovedisplayshortskip}{4pt}
 \begin{align*}
&\min ~~~~ |{\bf y}|_1 = \sum_{i = 1}^{n}y_i \\
\emph{s.t.}& ~(1)~ \bf \exists x, y = V_d^{(k)}x 
 ~(2)~ {\bf y }\geq 0,~(3)~\mathbf{s}^T\mathbf{y} \ge 1.
\end{align*}

Constraint (1), which requires $\bf{y}$ be in the span of $\bf V_d^{(k)}$,
could be rewritten as
$[\bf V_d^{(k)}, -I ]  \left[ \begin{array}{c}   \bf{x} \\  \bf{y}  \end{array}  \right] = \bf{0}.$
Constraint (2) requires $\bf{y}\geq 0$ where $y_i$ indicates the likelihood that vertex $i$ belongs to the target community.
Constraint (3) enforces the seeds be in the support of sparse vector $\bf{y}$ where $s$ is an indicator vector for the seed set.

Solve the above linear programming problem $LP_1$, and sort vertices according to their random walk probability scores in $\mathbf{y}$ in decreasing order.
We can then select the top $\hat n$ vertices with the highest probabilities as the resulting community. One could simply select the top $|C|$ vertices if the size of
the target community $C$ is known. Otherwise,
we use heuristics provided in Section \ref{sec:initialCommunity} to determine $\hat n$ automatically.

\subsection{An Alternative Definition of the Local Spectral Subspace}

An alternative approach to define the local spectral subspace is to adopt the normalized adjacency matrix $\mathbf{N_{sym}}$ instead of $\mathbf{N_{rw}}$.  
As $\mathbf{N_{sym}}$ is symmetric, 
the two matrices share the same set of real eigenvalues, with the eigenvectors of $\mathbf{N_{sym}}$ scaled by $\mathbf{D^{1/2}}$.
Note that $\mathbf{(N_{sym}^T)^k = D^{-1/2} (N_{rw}^T)^k D^{1/2}},$
$$\mathbf{(N_{sym}^T)^k D^{-1/2} p^{(1)} = D^{-1/2} (N_{rw}^T)^k p^{(1)}}$$
for an initial probability $\mathbf{p^{(1)}}$.
Thus the sequence of vectors generated with the random walk matrix is closely related to
the sequence generated by the symmetrized matrix.

We know $\mathbf{\pi}$ where $\pi_i = d_i/2m$ indicates the stationary probability on the transition matrix $\mathbf{N_{rw}}$,
i.e. $\mathbf{N_{rw}^T \pi} = \mathbf{\pi}$. As
$$\mathbf{N_{sym}^T D^{-1/2}\pi = D^{-1/2} N_{rw}^T \pi = D^{-1/2}\pi},$$
in the alternative definition of local spectral subspace,
we defined another random walk by $\mathbf{p^{t+1}} = norm(\mathbf{N_{sym}^T p^{t}})$,
which converges to the stationary distribution $\mathbf{\pi'}$ where $\pi_i' = \frac{\sqrt{d_i}}{\sum \sqrt{d_i}}$.
Further discussion on the two matrices is provided in Section \ref{sec:experiments_LOSP}.

For a simple example with two overlapping cliques, as shown in Fig. \ref{smallGraph1},
using the three red vertices as the seeds, we get the local spectral subspace calculated on $\mathbf{N_{rw}^T}$ for $k = 3$.
Finding the minimum $\ell_{1}$ norm sparse vector results in the left clique.
The same result holds for $\mathbf{N_{sym}}$.

\begin{center}
\begin{figure}
~\includegraphics[width=1.5in]{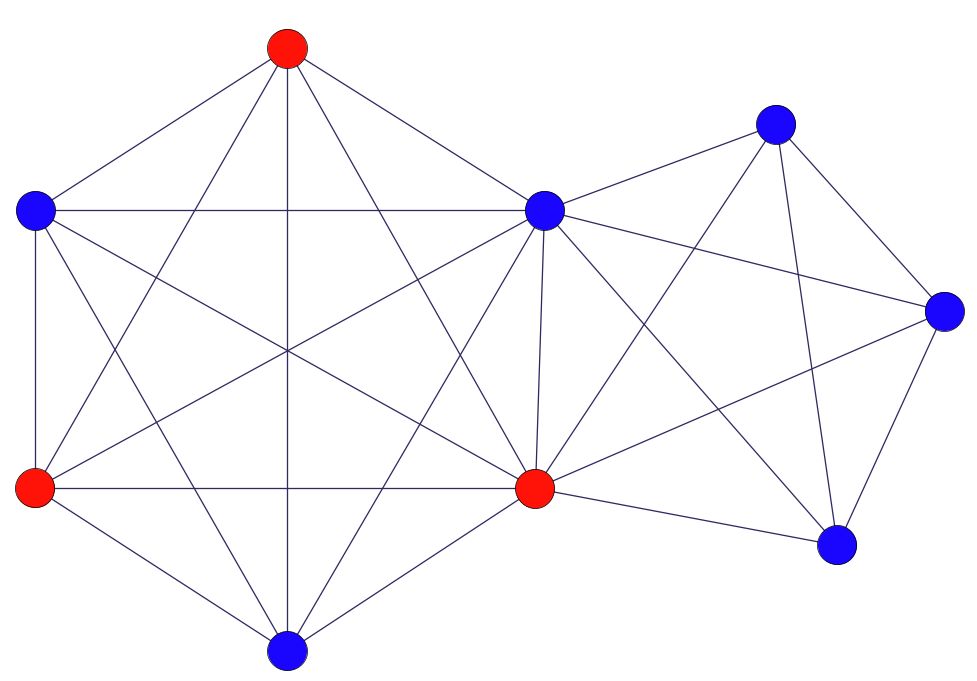}
\begin{tikzpicture}
\draw (-0.8,-0.1)[left] node{\parbox{2in}
{\tiny{\[\left[\begin{array}{ccc}
-0.29& -0.35\\
-0.29& -0.35\\
-0.29& -0.35\\
-0.29& -0.35\\
-0.48& ~~0.07\\
-0.48& ~~0.07\\
-0.26& ~~0.41\\
-0.26& ~~0.41\\
-0.26& ~~0.41\\
\end{array}\right]\]}}};
\end{tikzpicture}
\caption {An example network and its local spectral subspace of the seeds.}
\label{smallGraph1}
\end{figure}
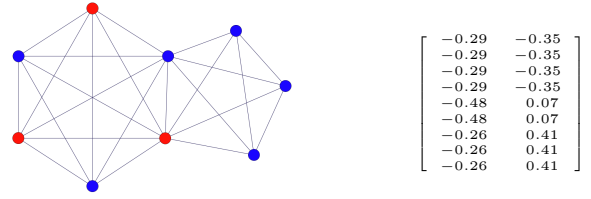
\end{center}

\section{The Proposed Algorithm}
\label{sec:Algorithm}
The whole local spectral (\emph{LOSP}) algorithm includes several procedures where the method of finding small communities on the local spectral subspace is a key subroutine.

\subsection{Sampling}

 The 90-percentile effective diameter $\mathcal{D}_{90\%}$ in Table \ref{real_data_summary} shows that for most vertices there is a small world phenomenon in which the distance between two randomly chosen vertices grows proportionally to $log(n)$\cite{SmallWorld1998}\footnote{One exception is on Amazon, which may be due to the peak at 4 in the degree distribution.}. This indicates that a community which is internally cohesive would have a smaller diameter than $log(n)$.

Rather than working on a large network with millions to billions of vertices,
we apply a local sampling method by breadth-first-search (BFS) to get a small subgraph that includes most of the neighborhood vertices.

\setlength{\belowdisplayskip}{4pt} \setlength{\belowdisplayshortskip}{4pt}
\setlength{\abovedisplayskip}{4pt} \setlength{\abovedisplayshortskip}{4pt}
\begin{itemize}
  \item For each vertex in the seed set $S$, do a 2-step BFS (3-step BFS on Amazon due to its larger $\mathcal{D}_{90\%}$).
  \item Consider the set of the frontier vertices, and remove vertices whose outdegree is greater than 1000 if there are any. Sort the remainder vertices in decreasing order according to their inward ratio (the fraction of inward edges to the BFS subgraph), and remove vertices after the first 1000.
  \item Union the BFS subgraphs obtained from each seed.
\end{itemize}

For a set $S$ with three random seeds, we generally yield a sampling subgraph of size thousands.
The coverage ratio, defined as the fraction of ground truth vertices covered by the sampled network, is very high at $90\%$ to $99\%$ except on Orkut whose ground truth communities are much larger.



The sampling method extracts a subnetwork in $\Theta(d_{avg}^r|S|)$ time where $r$ is the expanding level of the BFS procedure.
 depending only on the average degrees of the component and the number of initial seeds.
As follow-up procedures are only applied to this sampled network, the complexity is reduced
in time that is a function of the size
of the sampled network rather than that of the entire large network.
Note that we also effectively reduce the spatial complexity by only loading a small portion of the network into the memory.

\subsection{Strengthening the Initial Seed Set}


It is observed that larger seed set would lead to higher quality results\cite{IsabelKDD14}.
However, in many situations, we only know very few seed members.
To compensate for the shortage of available seeds, 
we provide a preprocessing procedure to enlarge the initial seed set before feeding the seeds into the local community detection algorithm.

Find a shortest path for each pair of vertices, and
add vertices on the path to the initial seed set if the length is no greater than a small number $l$.
We experimented with several values of $l$ including $l=1$ where no vertices will be added, and find that $l = 3$ yields the best results for Amazon and 4 for all other datasets. 

 The intuition behind this idea is that any two seeds in the same community must be related for some reason. They connect with each other either via a direct link or via some other intermediate vertices. In the latter case, those intermediate vertices bridging the seeds are very likely to be in the target community because they serve as the relational ``relay" in order for the seeds to be in the same community. The shorter the path is, the more likely the bridging vertices are in the same community.

\subsection{Initial Membership Identification}
\label{sec:initialCommunity}
In the key procedure of \emph{LOSP}, we feed the strengthened initial seed set $S$ to the linear programming problem $LP_1$ to find an initial community.
We need to determine the dimension of the local spectral subspace and the steps of the random walks for the subspace iteration, and also the size of the community.

\textbf{Dimension of the subspace \bf{\emph{d}}.} The dimension of the local spectral subspace is related to the number of local structures around the seeds.
We calculate the overlapping membership $om$, the number of communities a vertex is in, for vertices belonging to at least one ground truth, and found that $om$ is low at 1.2 for DBLP and around 1.5 for other datasets. As there may be some other community structures out of the annotation, we choose $d = 3$ assuming there are three local structures on average.

\textbf{Steps of the random walk \bf{\emph{k}}.}
 The number of steps of the random walk is related to size of the target community and its conductance. A larger or sparser community usually needs more steps to spread out the information, and a lower conductance serves as a better bottleneck to avoid probability leaking out. Experiments show that the detection accuracy plateaus as $k$ increases and $k = 3$ yields the full potential.

%

\textbf{Boundary of the local community.}
As the global minimum of scoring functions like conductance might produce ``cavemen-type" communities\cite{Kang2011ICDM},
we use a local optimum to decide the size of the community.
 For each set $S_i$ with the $i$ vertices having the highest probability scores, we evaluate its quality using a candidate community scoring function $f$.
The truncation at the first local optimum of $f(S_i)$ corresponds to the extracted community. 

In minimizing a scoring function such as conductance,
we increase the index $i$ in the sorted $\bf{y}$ to find the first minimum.
Since the function does not smoothly decrease to the minimum, we search for indices $n_0$ and $\hat n$ ($n_0 < \hat n$),
such that the scoring function $f$ starts increasing at $\hat n$ and the drop from $f(n_0)$ to $f(\hat n)$ satisfies $f(n_0) \geq \gamma f(\hat n)$.
Experiments with several values show that $\gamma = 1.7$ yields good results across all the datasets.


\subsection{Reseeding} 

After the initial membership identification, we further improve the detection quality by iteratively augmenting the seed set with the top elements in the sorted $\bf{y}$ and use the enlarged seed set to find the community again.
The intuition is that the top elements with high probabilities are likely to be in the target community,
 and augmenting the initial seed set should uncover a more accurate target community.

Let $S_0 = S$, and consider the top $|S|+\delta \cdot t$ candidate seeds at iteration $t$.
 Union this augmented seed set with $S$ in case some initial seeds are excluded from the current top\footnote{We expand the candidates by $\delta = 5$ at each iteration in the experiments.}.
Define the weight of each initial seed in $S$ to be $w_1 = 1/|S|$,
and the weight of each augmented seed in $S_t/S_0$ to be $w_2 = 0.5 w_1$.
Then feed the expanded seed set to the modified linear programming problem $LP_2$:
 \setlength{\belowdisplayskip}{4pt} \setlength{\belowdisplayshortskip}{4pt}
\setlength{\abovedisplayskip}{4pt} \setlength{\abovedisplayshortskip}{4pt}
 \begin{align*}
\min &~~~~ |{\bf y}|_1 = \sum_{i = 1}^{n}y_i \\
\emph{s.t.}& ~~~(1)~[\bf V_d^{(k)}, -I ]  \left[ \begin{array}{c}   \bf{x} \\  \bf{y}  \end{array}  \right] = \bf{0} \\
&~~~(2)~\quad {\bf y }\geq 0,\\
&~~~(3)~\mathbf{s}^T\mathbf{y} \ge 1,\\
&~~~(4)~\mathbf{s_t}^T\mathbf{y} \ge 1 + w_2*(|S_t| - |S|),
\end{align*} %
where $\mathbf{s_t}\in \{0,1\}^n$ is a binary indicator vector for the current seed set $S_t$. We halve the weight of the expanded seeds such that the initial members play a key role for the identification.

To complete the reseeding process, we track the value of the scoring function on the extracted community during the iterations, and stop the reseeding process when the community quality starts to decline.

\section{Multiple Membership Identification}
\label{sec:FindAllComm}
In real world networks, vertices tend to belong to multiple communities  simultaneously.
We further wish to answer how many communities is a single individual in and how do we identify the communities.

 Regard the individual seed $s$ as an ``ego", and temporarily remove the seed from its ego network to get connected components sorted by their sizes in decreasing order, namely $S_1$, $S_2$, ..., $S_q$. Each $\{s\} \cup S_i$ is regarded as a candidate initial seed set.
 We then iteratively start from each initial seed set, remove edges connecting $s$ to other ego neighbors, and identify the corresponding community by using \emph{LOSP} as a subroutine.
 Seed sets totally contained in previous communities are not processed.

  A crucial step is that we cut edges connecting $s$ to other candidate seed sets so as to weaken the interference of different seed sets.
  If a candidate seed set still has very strong connections to the current target structure, it will be totally covered by the extracted community and then be removed from the candidate seed list.
  In this way, we find a set of local overlapping communities containing the single seed. The first is the community the seed is mostly attached to.


\section{Experimental Results}
\label{sec:experiments_Comparison}
We implement \emph{LOSP} in Matlab ~\footnote{https://github.com/KunHe2015/LOSP/}
and compare our results with several state-of-the-art local as well as global algorithms.
The experiment was on a server with 2 Intel Xeon processors at 2.0GHz and 64GB memory. 
For each of the real world datasets, we randomly locate 100 or 500 out of the 5000 labeled ground truth communities,
 and randomly pick one or three exemplary seeds from each target community.
The mean and standard deviation of F1 scores over all trials are used for the evaluation and comparison.

\subsection{Evaluation on LOSP}
\label{sec:experiments_LOSP}

We randomly pick 500 ground truth communities for the evaluation of different variations and procedures of \emph{LOSP}.
Throughout this subsection, unless otherwise pointed out,
 communities are truncated by the size of the target community to remove the impact of other factors
 (except the evaluation for different community scoring functions),
and communities are extracted by starting from three random seeds.

\textbf{Normalized Matrices.}
$\mathbf{N_{rw}}$ and $\mathbf{N_{sym}}$ are related to the normalized graph Laplacian
$\mathbf{L_{rw}} = \mathbf{I} - \mathbf{N_{rw}}$ and $\mathbf{L_{sym}} = \mathbf{I} - \mathbf{N_{sym}}$.
In the classical spectral method, for which graph Laplacians are used to compute the eigenvectors before the clustering,
if the degrees are very broadly distributed, which is usually the case in real world networks,
it is preferable to use normalized rather than unnormalized spectral clustering, and in the normalized case
 to use $\mathbf{L_{rw}}$ rather than $\mathbf{L_{sym}}$ \cite{Spectral2007}.

Table \ref{comparison_Matrix} shows the average F1 scores using the two different local spectral subspaces defined by the two matrices $\mathbf{N_{rw}}$ and $\mathbf{N_{sym}}$.
$\mathbf{N_{rw}}$ also performs better than $\mathbf{N_{sym}}$ on average.


\begin{table}[htbp]
\begin{center}
\begin{tabular}{l c c c c c c }
\hline
                   &\bf{{Amazon}}&\bf{{DBLP}}&\bf{{LiveJ}} &\bf{{YouTube}}&\bf{{Orkut}}&\bf{{Avg.}} \\ \hline
$\mathbf{N_{rw}}$  & 0.938&	0.911&	0.726&	0.591&	0.261&	0.685 \\
$\mathbf{N_{sym}}$ & 0.920 & 0.845 & 0.751 & 0.531 & 0.277 & 0.659 \\\hline 
\end{tabular}
\end{center}
\caption{\bf{Evaluation of LOSP with different Laplacian matrices.}}
\label{comparison_Matrix} %
\end{table}

\begin{figure}[t]
\begin{center} \includegraphics[width=2.9in]{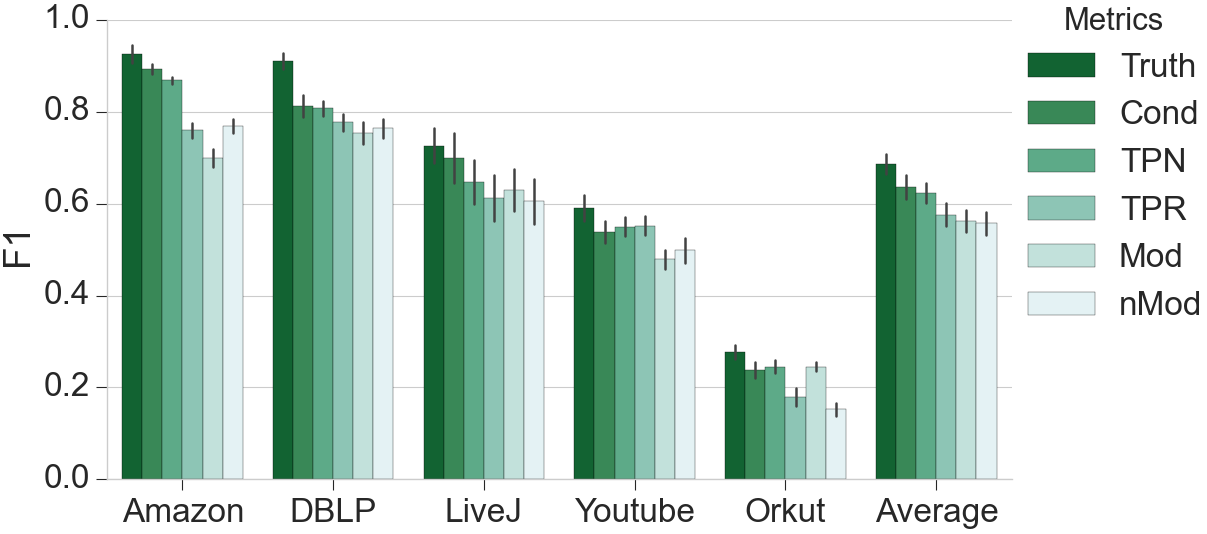}
\vspace{-0.5em}
\caption{\bf Evaluation on different community scoring functions.} 
\label{fig:Evaluations_Metrics}
\end{center}
\end{figure}

\textbf{Community Scoring Functions.}
How to automatically determine the community boundary is crucial for the quality of the community.
\emph{LOSP} searches the first local optimum of a scoring function
on the sorted probability $\mathbf{y}$ to do the truncation.
Fig. \ref{fig:Evaluations_Metrics} shows
the average F1 scores and standard deviations of the resulting communities for different scoring functions.
\emph{LOSP} is robust in revealing communities resembling the ground truth,
as there is a low variance on different scoring functions.

TPN consistently outperforms other metrics by producing the best results.
This may due to the fact that TPN quantifies the number of triads each vertex is in to define close-knit communities.
Also, though TPN only explicitly considers the internal connectivity,
it implies a sparse connectivity to the remainder of the network.
If an external vertex has more connections to a community compared with the internal members,
 adding the vertex to the community will definitely increase the TPN score.

Conductance is almost as good as TPN.
Compared with modularity, the normalized modularity outputs higher accuracy results on Amazon and LiveJ,
but lower accuracy on the other three datasets.

\begin{figure}[ht]
\begin{center} \includegraphics[width=3.2in]{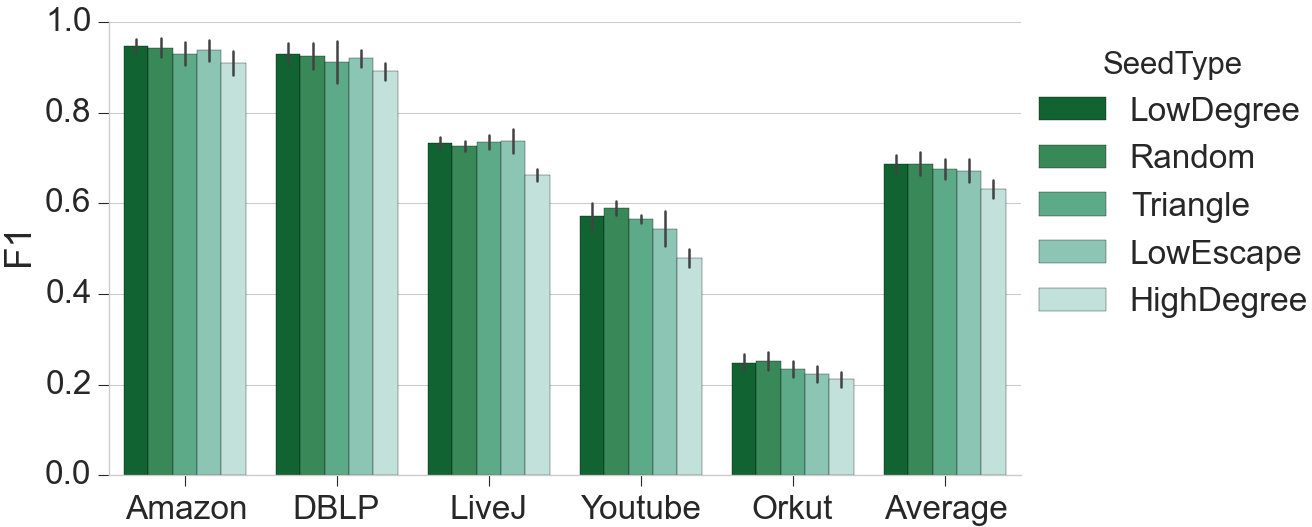}
\vspace{-0.5em}
\caption{\bf Evaluation on different seed structure.} 
\label{fig:seedStructure}
\end{center}
\end{figure}

Note that we work on the sampled subgraph when calculating the scoring functions for candidate communities.
Therefore, the modularity and other metrics are ``local", and in this way \emph{LOSP} alleviates the negative impact of the resolution limit.

\textbf{Seed Structure.}
To understand how the structure of the initial seed set affects the accuracy of the resulting community,
we evaluate the performance of five different seeding strategies, including two new seed structures,
high triangle participation seeds and low escape seeds.
\setlength{\belowdisplayskip}{4pt} \setlength{\belowdisplayshortskip}{4pt}
\setlength{\abovedisplayskip}{4pt} \setlength{\abovedisplayshortskip}{4pt}
\begin{itemize}
\item {\bf Random seeds}: a natural way is to randomly pick $|S|$ vertices from the target community $C$.
\item {\bf{High degree or low degree seeds}}: randomly pick $|S|$ vertices with degree ranked in the top or bottom
one third among the degree of all vertices in ${C}$.
\item {\bf High triangle participation seeds}: sort vertices according to the number of triangles they belong to in $C$,
and randomly pick $|S|$ vertices ranked in the top one third.
\item {\bf Low escape seeds}: sort vertices basing on the probability reserved
after a few steps of random walk starting from each vertex,
and randomly pick $|S|$ vertices in the top one third.
\end{itemize}

Fig. \ref{fig:seedStructure} shows that low degree, random, triangle (high triangle participation) and low escape yield almost
 the same accuracy on average.
Due to the power law distribution of the degrees of the vertices,
most of the vertices are of low degree and we rarely select high degree vertices when picking randomly,
 which makes the seeding of low degree and of random almost the same.
 Low degree seeds spread out the probabilities slowly and better preserve the information of the local structure.
High triangle participation seeds and low escape seeds follow another philosophy in that they choose seeds more
cohesive to the target community, resulting in high quality output.

High degree seeds are inferior to the previous four seed structures.
The intuition behind this phenomenon is that the probabilities
spread out very quickly via short random walks from these popular individuals, causing less accuracy.
In \cite{IsabelKDD14}, the authors evaluated different PageRank based seed set expansion algorithms,
 and conclude that random seeds always outperform high degree seeds in real networks across domains,
  which is consistent with our results.
  We further observe that random seeds is essentially low degree seeds for real networks following the power law distribution.
  Also, high degree vertices behave as the ``hubs" while low degree vertices has a higher loyalty to the target community.

We define the cohesive degree of a seed $s$ to a community that $s$ is in to quantify how cohesive or active this seed is in this community.

\vspace{-0.5em}
\begin{definition}
\textbf{Cohesive Degree}. The cohesive degree of a seed member $s$ to a community $C_k$ is the fraction of its internal connections to the fraction of its external connections.
$$ Coh(s,C_k) = \frac{e_{sk}/n_k}{e_{s\bar{k}}/n }$$
where $e_{sk}$, $e_{s\bar{k}}$ are the number of edges connecting $s$ to $C_k$ and $\overline{C_k}$.
\end{definition}
\vspace{-0.5em}

\begin{figure*}[t!]
  \subfigure[TPN]{
    \begin{minipage}[b]{0.46\textwidth}
      \centering
      \includegraphics[width=3.2in]{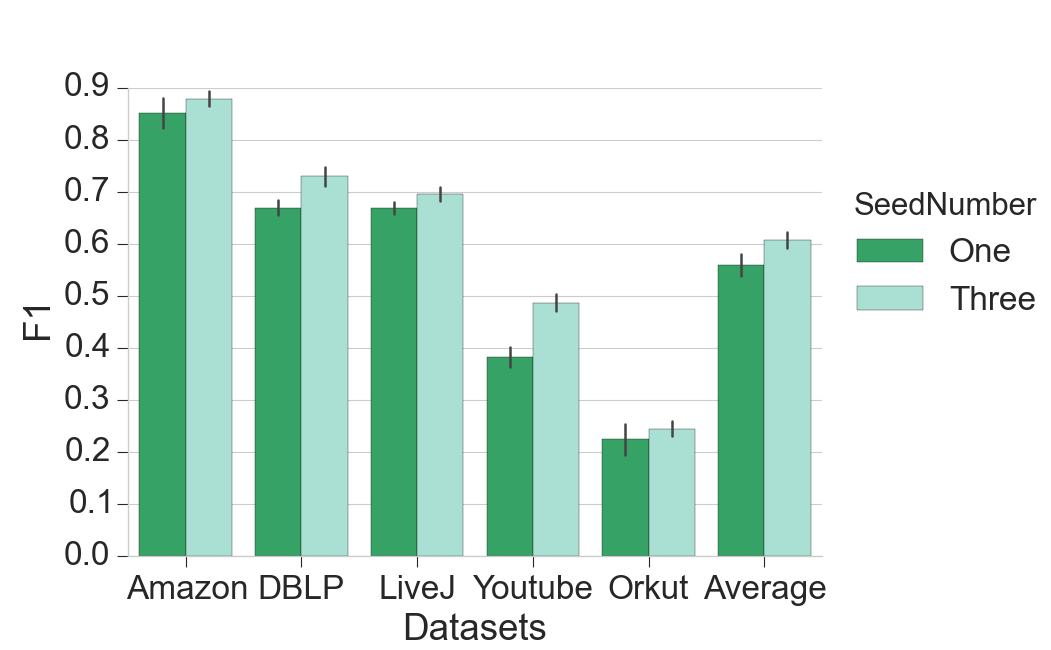}
    \end{minipage}}
  \subfigure[Cond]{
    \begin{minipage}[b]{0.46\textwidth}
      \centering
      \includegraphics[width=3.2in]{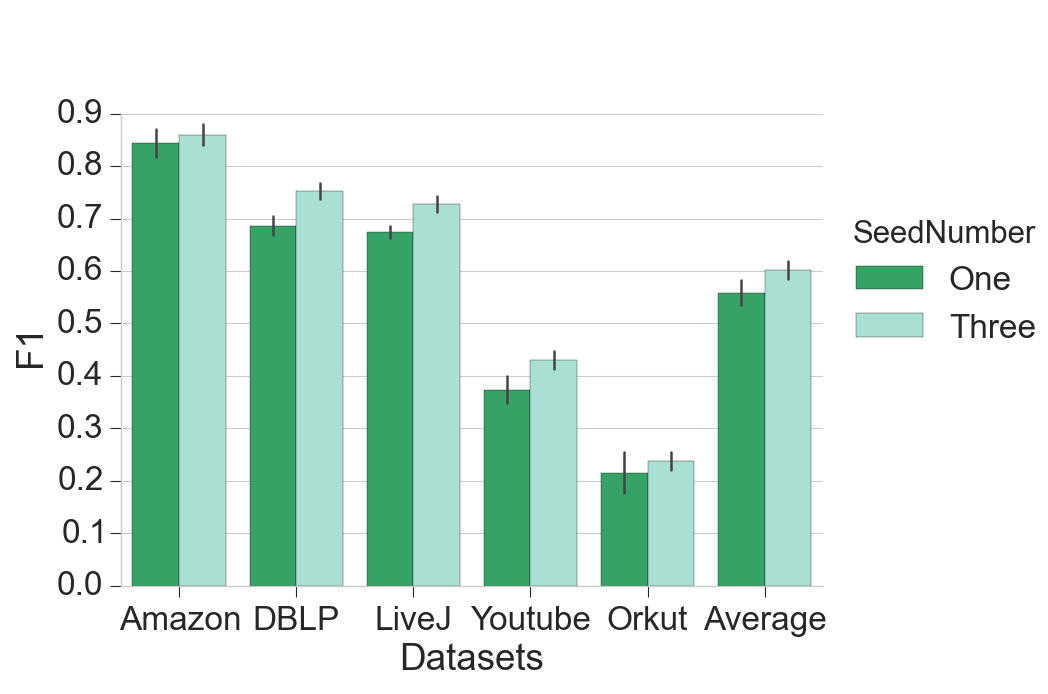}
    \end{minipage}}
    \vspace{-1em}
 \caption{Evaluation on the different seed quantity.} 
  \label{fig:oneSeedvsThreeSeed} 
\end{figure*}

\textbf{Seed Set Size.}
Fig. \ref{fig:oneSeedvsThreeSeed} shows
the average F1 scores of the communities by starting from one random seed or three random seeds.
To save space, we only show TPN and Conductance as the scoring functions.
\emph{LOSP} is robust in that it outputs high F1 score communities even when starting from a single seed.

\begin{figure*}[t!]
  \subfigure[TPN]{
    \begin{minipage}[b]{0.46\textwidth}
      \centering
      \includegraphics[width=3.2in]{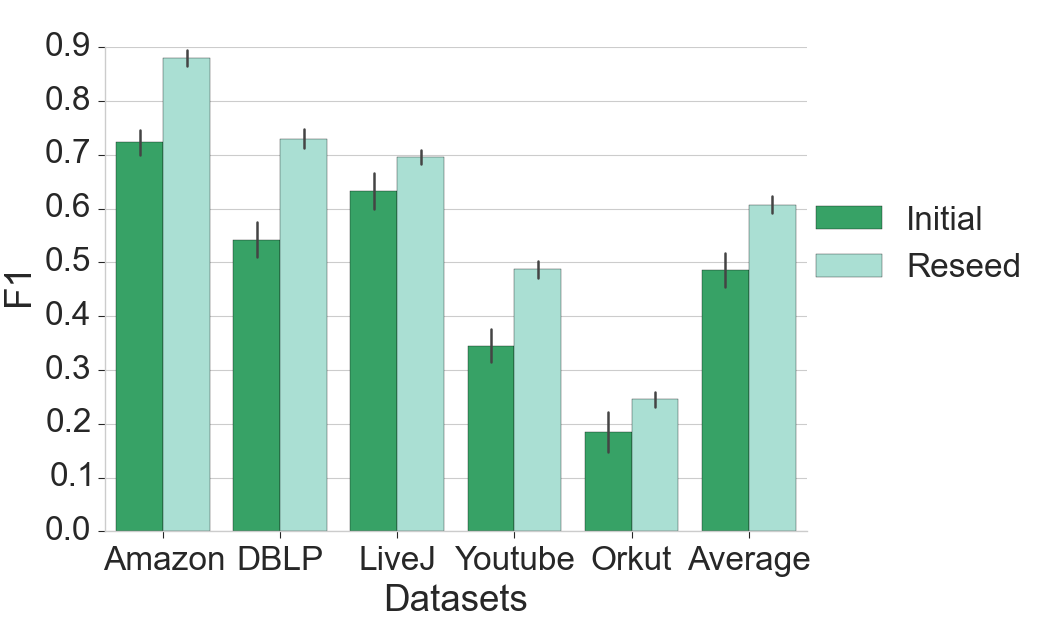}
    \end{minipage}}
  \subfigure[Cond]{
    \begin{minipage}[b]{0.46\textwidth}
      \centering
      \includegraphics[width=3.2in]{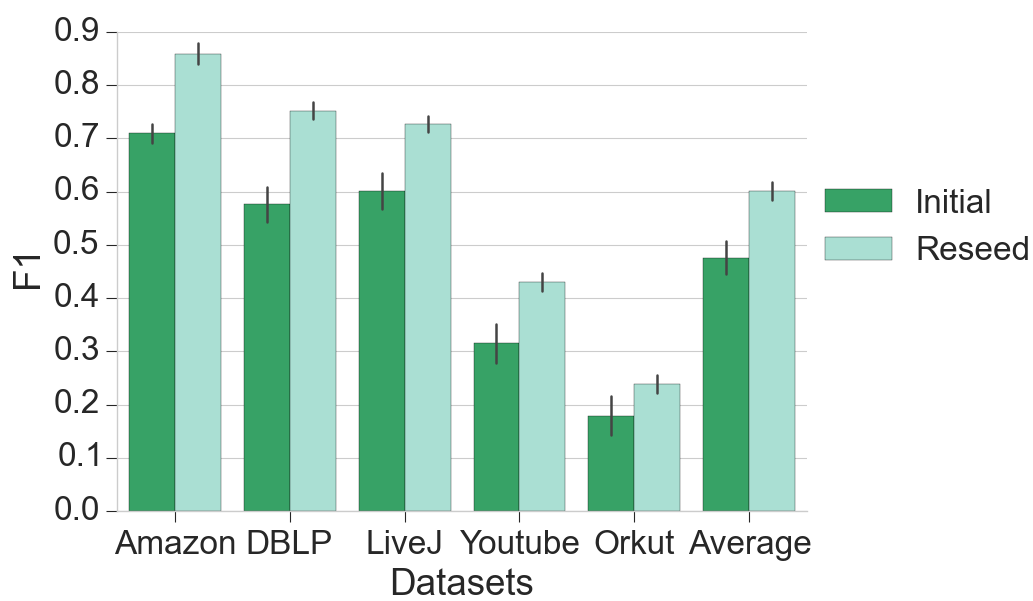}
    \end{minipage}}
    \vspace{-1em}
  \caption{Evaluation on seeding and reseeding.} 
  \label{fig:InitialvsReseeding} 
\end{figure*}

\textbf{Seeding versus Reseeding.}
Fig. \ref{fig:InitialvsReseeding} shows the average F1 scores for the initial communities and the final communities after reseeding by starting from three exemplary seeds.
When using TPN or conductance as the scoring function, the F1 scores increase considerably, by 0.12 or 0.13 on average.
By comparison, the running time with reseeding is almost four to seven times the initial. The reseeding process is worthwhile as \emph{LOSP} is very quick in seconds for large networks with  millions of vertices.

\subsection{Comparisons with other algorithms}

We randomly pick 500 ground truth communities for the final results of \emph{LOSP},
and compare with several state-of-the-art local and global community detection algorithms.
The results of LOSP\_{Truth} is obtained by using the size of the ground truth community for the truncation,
which could be approximately regarded as the upper bound of \emph{LOSP} using scoring functions to decide the community boundary.
The results by applying the best two scoring functions, TPN and conductance, are reported.

\textbf{Comparisons with local algorithms.} To compare with the local methods, we randomly pick one seed from each located community
 for the local structure identification,
and compare the average F1 scores by running other local algorithms under the same condition.
For the Heat Kernel algorithm\footnote{\url{https://gist.github.com/dgleich/cf170a226aa848240cf4}} \cite{kloster2014heat},
we choose its best variation hk-relax for the comparison.
The authors of Heat Kernel compared their results with pprpush\cite{PageRank2006},
the PageRank Nibble local clustering algorithm, and Heat Kernel far outperforms pprpush by almost doubling the F1 scores.
Table \ref{comparison_local} shows that \emph{LOSP} apparently outperforms Heat Kernel, especially on DBLP and YouTube.
We then randomly pick three seeds from each target community, and compare \emph{LOSP} with LEMON~\cite{li2015uncovering} that reports result using conductance as scoring function and starts from three random seeds. Table \ref{comparison_local} shows that \emph{LOSP} apparently outperforms LEMON.

\begin{table}[htbp]
\begin{center}
\scalebox{1}{
\begin{tabular}{l c c c c c c }
\hline
\bf{{Algorithm}}&\bf{{Amazon}}&\bf{{DBLP}}&\bf{{LiveJ}} &\bf{{YouTube}}&\bf{{Orkut}}&\bf{{Avg.}} \\ \hline
\emph{\small{one seed:}}\\
\bf{{LOSP\_{Truth}}}&0.864 & 0.734 &	0.686 & 0.476 & 0.247 & 0.601 \\  
\bf{{LOSP\_{Cond}}}& 0.845&	0.691&	0.674&	0.413&	0.216&	0.568 \\
\bf{{LOSP\_{TPN}}}&  0.722&	0.686&	0.669&	0.406&	0.224&	0.542 \\
\bf{{HeatKernel}}& 0.712 &	0.378 &	0.553 &	0.098 &	0.320 &	0.412 \\ 
 \hline
\emph{\small{three seeds:}}\\
\bf{{LOSP\_{Truth}}}& 0.938&	0.911&	0.726&	0.591&	0.261&	0.685 \\  
\bf{{LOSP\_{Cond}}} & 0.893&	0.812&	0.699&	0.538&	0.234&	0.635 \\
\bf{{LOSP\_{TPN}}}  & 0.868&	0.807&	0.646&	0.550&	0.231&	0.620 \\
\bf{{LEMON\_{Cond}}}& 0.910&    0.525&    -  &  0.190&  0.170&    -  \\
\hline
\end{tabular}}
\end{center}
\caption{\bf{Comparison with local algorithms.}}
\label{comparison_local}
\end{table}

In another important work on local community detection, the authors \cite{IsabelKDD14}
report their recalls on Amazon and DBLP using $10\%$ of the random vertices in the community as the seeds.
They did not explicitly determine the community boundary,
and instead used a budget for predicting the size of the target community.
For reference, even if we choose a fairly large budget of 200 as the community size,
only a recall of 0.2 is achieved for both Amazon and DBLP.
Note that their recall is defined slightly differently, by subtracting the seed size from the denominator,
which increases the value by a factor of 1.11 on the standard recall.

\begin{figure*}[t!]
  \subfigure[Amazon]{
    \begin{minipage}[b]{0.18\textwidth}
      \centering
      \includegraphics[width=1in]{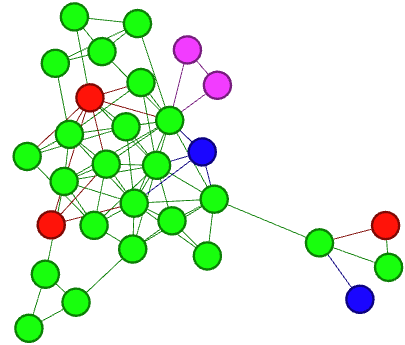}
    \end{minipage}}
  \subfigure[DBLP]{
    \begin{minipage}[b]{0.18\textwidth}
      \centering
      \includegraphics[width=1.3in]{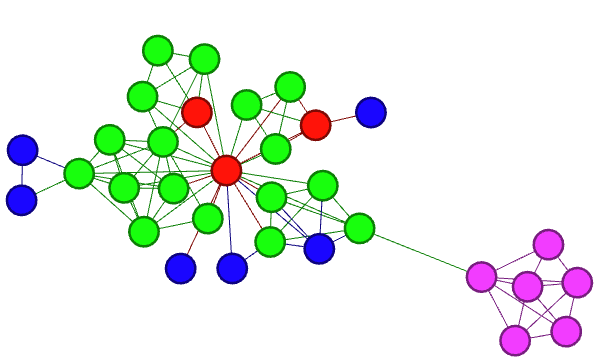}
    \end{minipage}}
\subfigure[LiveJ]{
    \begin{minipage}[b]{0.18\textwidth}
      \centering
      \includegraphics[width=1in]{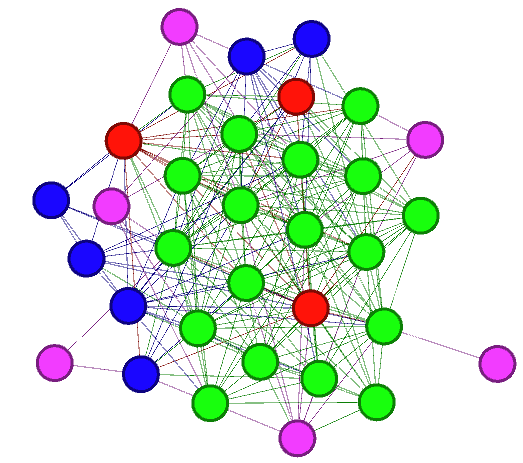}
    \end{minipage}}
\subfigure[Youtube]{
    \begin{minipage}[b]{0.18\textwidth}
      \centering
      \includegraphics[width=1.2in]{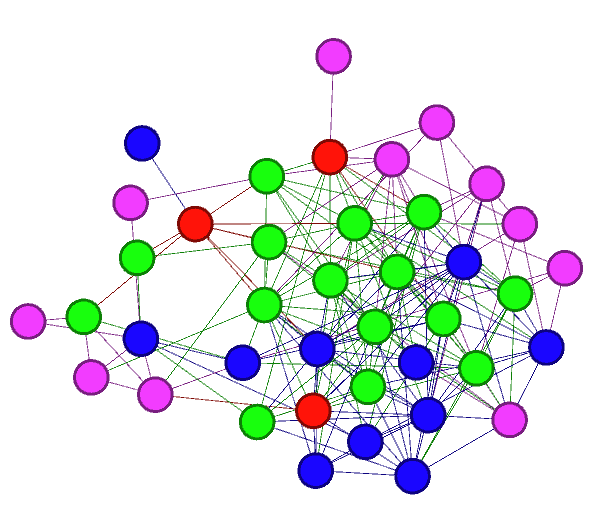}
    \end{minipage}}
\subfigure[Orkut]{
    \begin{minipage}[b]{0.18\textwidth}
      \centering
      \includegraphics[width=1.5in]{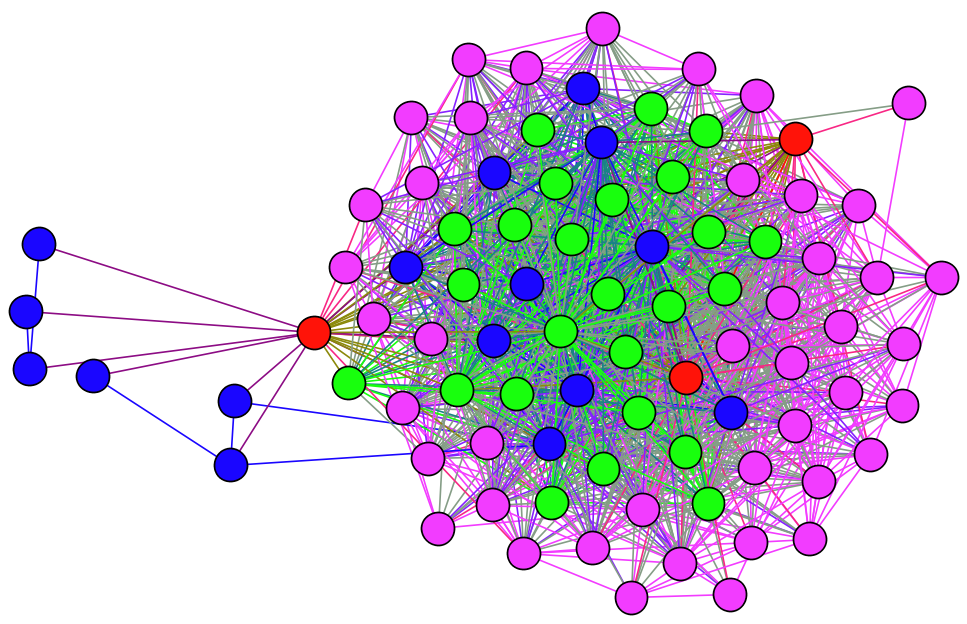}
    \end{minipage}}
\caption{An example of the detected communities compared with the ground truth for each dataset. The red vertices indicate the random seeds,
the green are the intersection of detected community and the ground truth, while the pink and blue are the remainders of the ground truth and additional detected vertices.}
  \label{fig:detected_vs_groundtruth}
\end{figure*}

\textbf{Comparisons with global algorithms.} To have a thorough understanding
 of the local structure mining method,
we also compare \emph{LOSP}
 with five state-of-the-art global community detection algorithms,
GCE\footnote{\url{https://sites.google.com/site/greedycliqueexpansion}} \cite{lee2010GCE},
SSE\footnote{\url{http://www.cs.utexas.edu/~joyce/codes/cikm2013/}}\cite{Whang2013SSE},
OSLOM\footnote{ \url{http://www.oslom.org/software.htm}}  \cite{lancichinetti2011OSLOM},
DEMON\footnote{\url{http://www.michelecoscia.com/?page_id=42}} \cite{coscia2012demon},
and LinkCommunity (LC)\footnote{\url{https://github.com/bagrow/linkcomm}} \cite{Ahn2010LinkCommunities}.
Table \ref{comparison_global} summarizes the average F1 scores and running times.
And Fig. \ref{fig:detected_vs_groundtruth} illustrates the accuracy of the detection for one randomly picked community of each dataset.

Almost all of the global algorithms fail to halt in 5 days on large networks like LiveJ and Orkut.
SSE was performed on a computer with a Xeon X5440 2.83GHz CPU and 32GB memory by the authors,
and they only reported results on relatively small networks Amazon and DBLP
\footnote{we use the result of Spread hubs which outputs the best results for SSE.}.
OSLOM and LC can achieve good results but do not scale well.
DEMON has consistently lower scores as it usually outputs communities larger
than the natural size of the ground truth.

By starting from only three random seeds, \emph{LOSP} obtains much better results than the baselines
and runs very fast in less than 5 seconds even on the largest dataset, Orkut.
Note that even for Orkut, \emph{LOSP} could also run on a personal computer with only 2GB memory.
By the efficient sampling method,
we only need to load a small portion of the local neighborhood around the seeds of interest,
causing a very low memory consumption.
Another strength of \emph{LOSP} is its high potential for parallelization
 as we could work on different seed set independently and simultaneously.

\begin{table*}[t]
\begin{center}
\begin{tabular}{l | r r | r r | r r | r r | r r | r r}
\hline
& \multicolumn{2}{c}{\bf{Amazon}} \vline& \multicolumn{2}{c}{\bf{DBLP}} \vline&\multicolumn{2}{c}{\bf{LiveJ}} \vline&\multicolumn{2}{c}{\bf{YouTube}}  \vline&\multicolumn{2}{c}{\bf{Orkut}} \vline&\multicolumn{2}{c}{\bf{Average}} \\
\bf{Algorithm}&  F1  & time& F1  & time  & F1  & time & F1  & time & F1  &  time  & F1  &  time \\
 \hline
\bf{{LOSP\_{Truth}}}& 0.938& 0.04s&	0.911& 0.38s&	0.726& 1.47s	&   0.591& 3.85s &	0.261& 4.74s &	0.685 & 2.10s\\  
\bf{{LOSP\_{Cond}}} & 0.893& 0.04s&	0.812& 0.38s&	0.699& 1.47s	&	0.538& 3.85s &	0.234& 4.74s &	0.635 & 2.10s\\
\bf{{LOSP\_{TPN}}}  & 0.868& 0.04s&	0.807& 0.38s&	0.646& 1.47s	&	0.550& 3.85s &	0.231& 4.74s &	0.620 & 2.10s\\\hline
\bf{GCE}          & 0.445   &    11s  & 0.524 &    18s &-  & -   &0.019 & $>$ 5d     & - & -& - &-\\
\bf{SSE}          & 0.490   &   1,260s& 0.181 & 1,152s &- & -   & -    & - 			& - & -& - &-\\
\bf{DEMON}        & 0.161   &  3,492s & 0.157 & 2,893s &- & -   &0.067 & 6,215s		& - & -& - &-\\
\bf{OSLOM}        & 0.766   &   3.24h & 0.570 &  6.05h &- &-    &0.080 & 158,992s 	& - & -& - &-\\
\bf{LC(C++)}      & 0.835   &     95s & 0.491 &   166s &- & -   &0.062 & 25,307s	&-  & -& - &-\\ \hline
\end{tabular}
\end{center} 
\caption{Comparison with global algorithms (\emph{LOSP} starts from three random seeds).}
\label{comparison_global}
\end{table*} 

\begin{figure*}[t!]
  \subfigure[]{
    \begin{minipage}[b]{0.15\textwidth}
      \centering
      \includegraphics[width=0.48in]{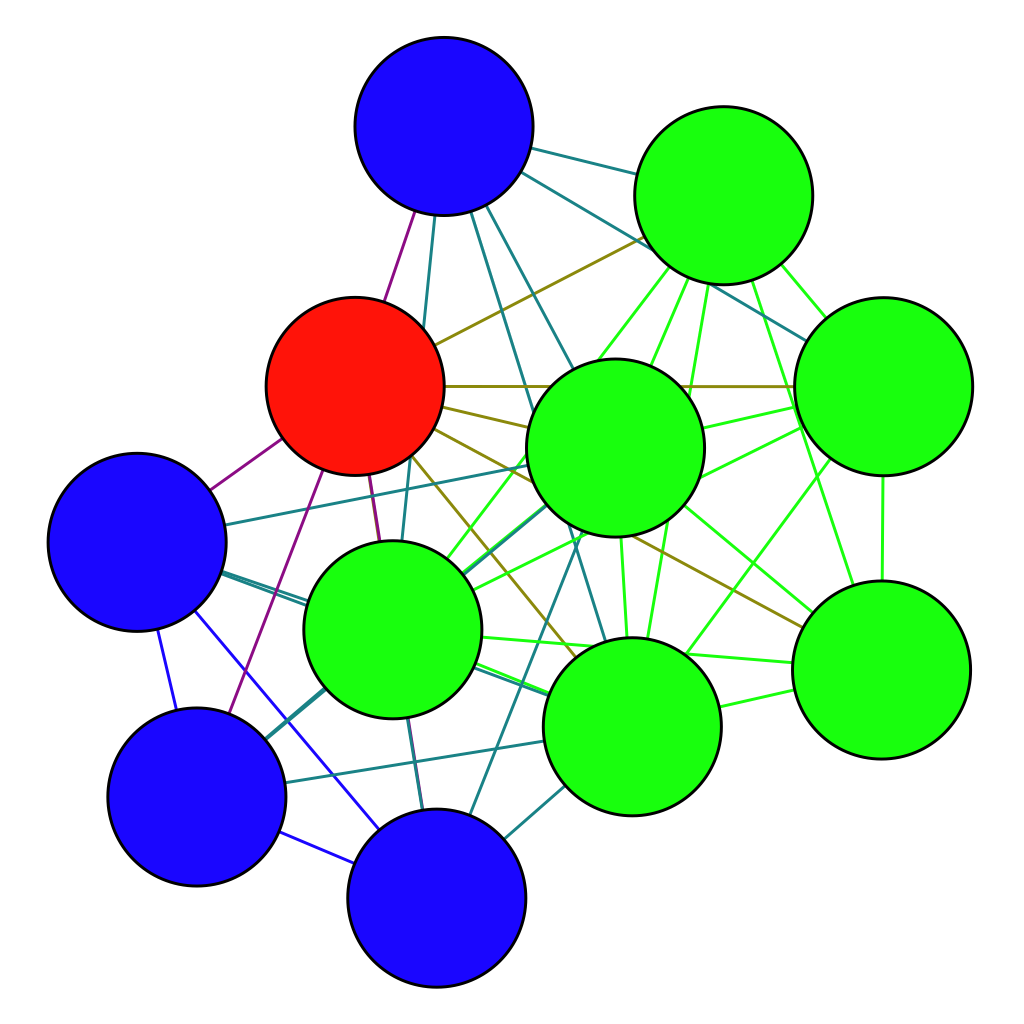}
    \end{minipage}}
  \subfigure[]{
    \begin{minipage}[b]{0.18\textwidth}
      \centering
      \includegraphics[width=0.72in]{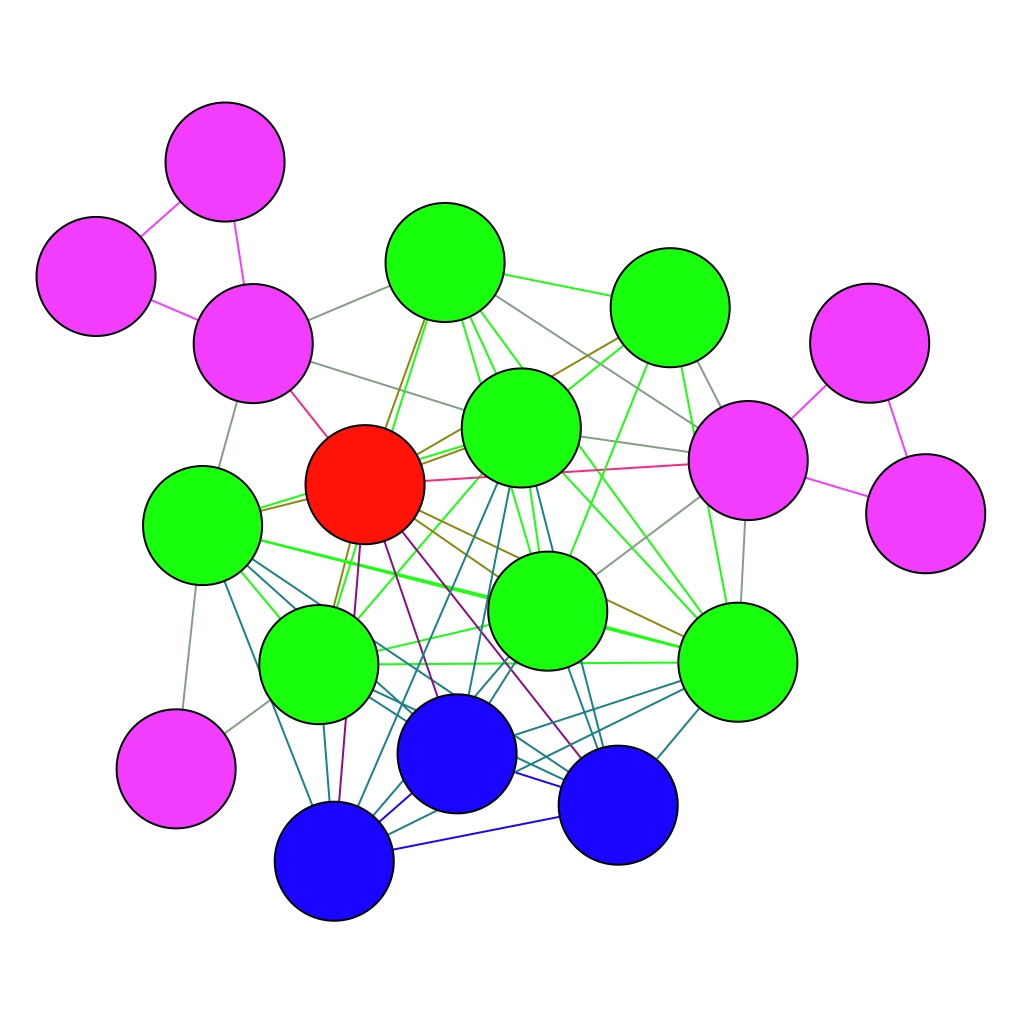} 
    \end{minipage}}
\subfigure[]{
    \begin{minipage}[b]{0.18\textwidth}
      \centering
      \includegraphics[width=0.8in]{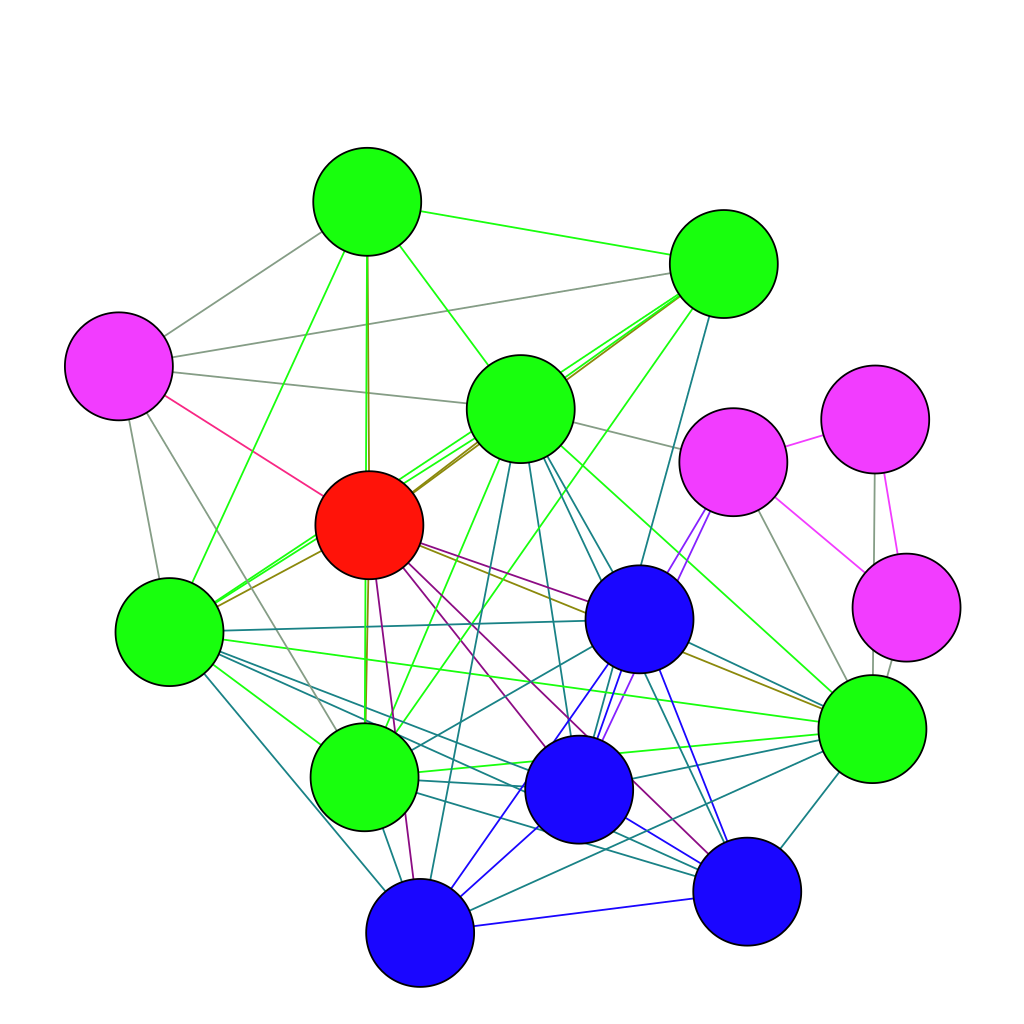}
    \end{minipage}}
\subfigure[]{
    \begin{minipage}[b]{0.18\textwidth}
      \centering
      \includegraphics[width=0.98in]{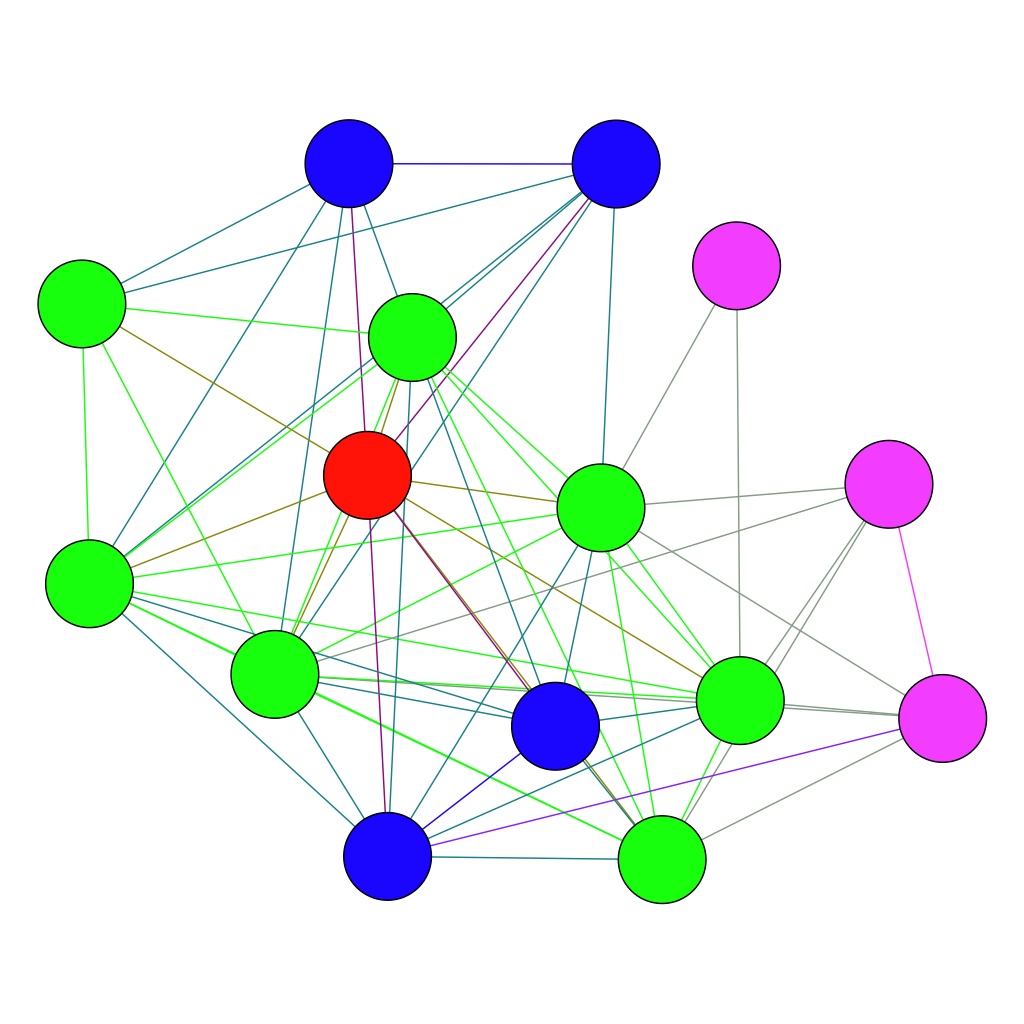} 
    \end{minipage}}
\subfigure[]{
    \begin{minipage}[b]{0.1\textwidth}
      \centering
      \includegraphics[width=0.28in]{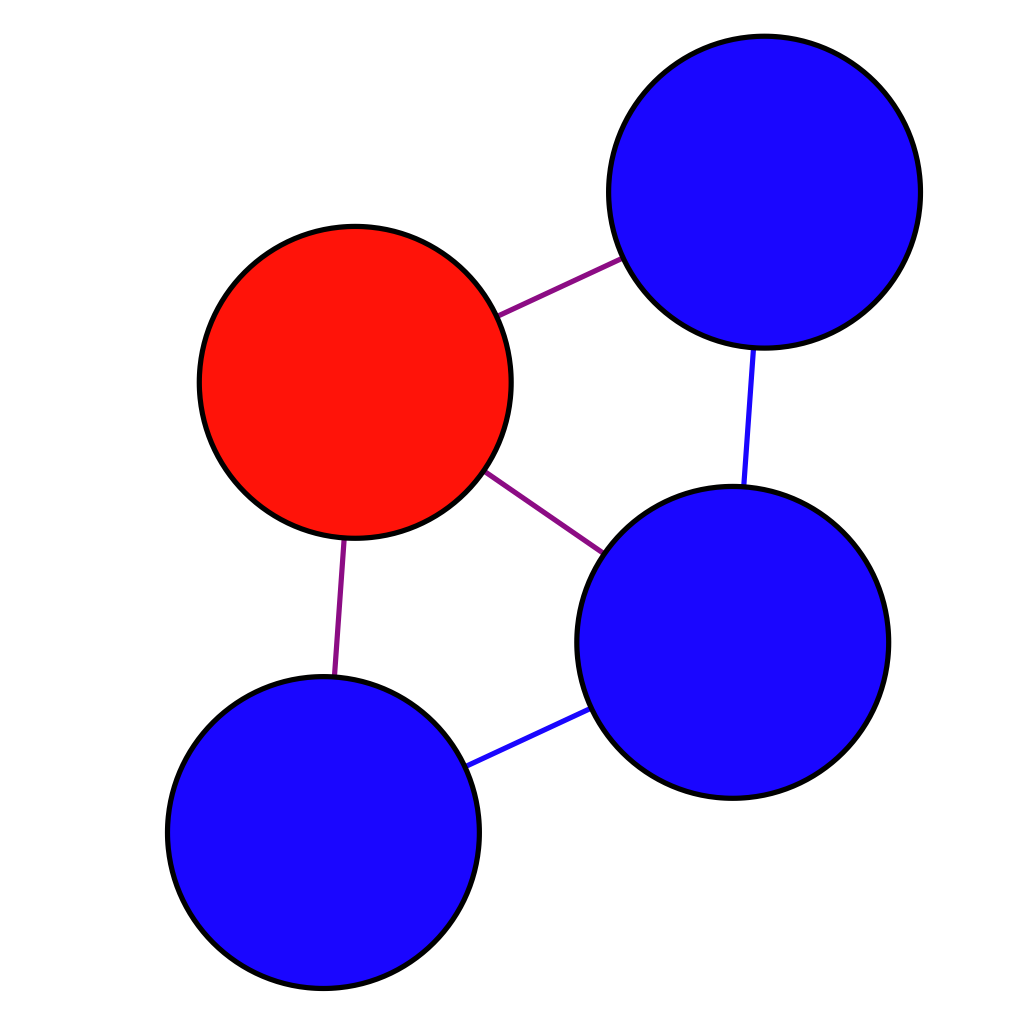} 
    \end{minipage}}
\subfigure[]{
    \begin{minipage}[b]{0.1\textwidth}
      \centering
      \includegraphics[width=0.28in]{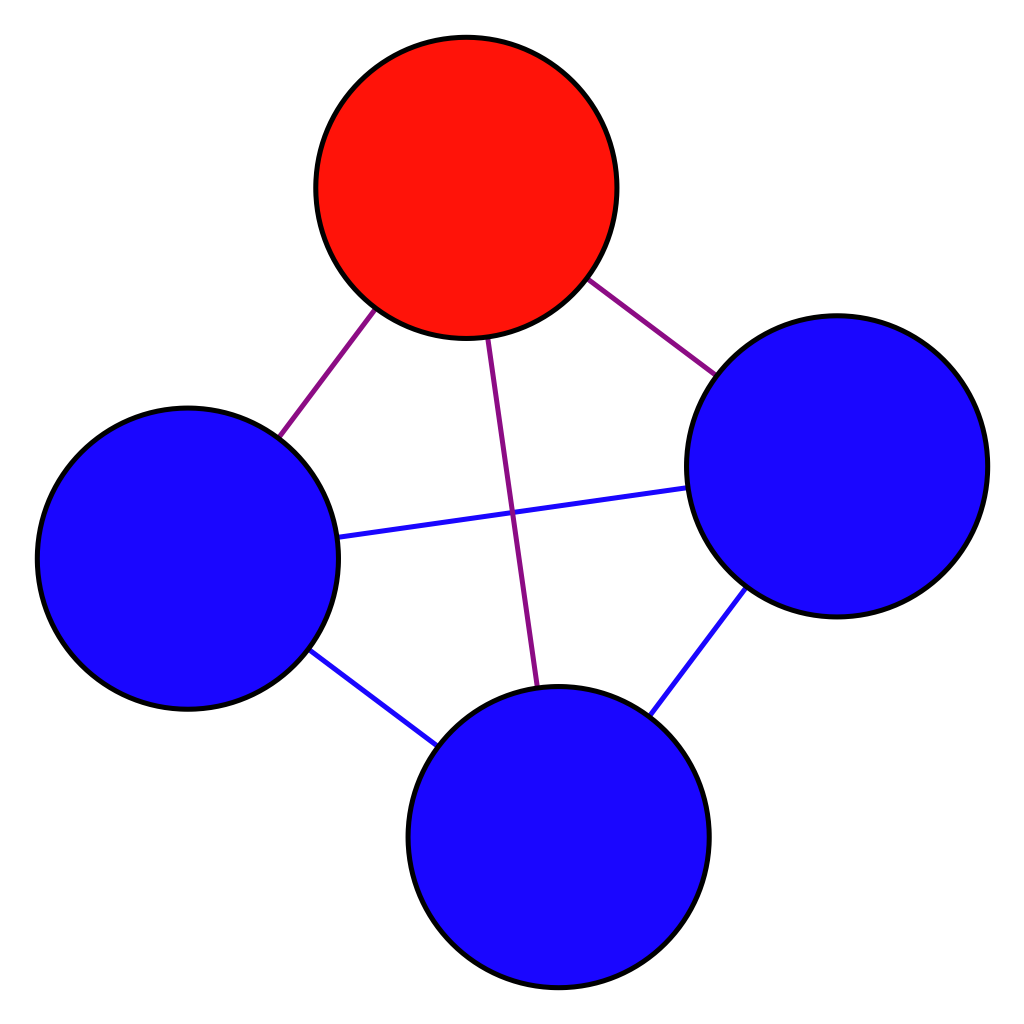} 
    \end{minipage}}
\caption{An example of all local memberships of a vertex picked from the DBLP network. The picked red vertex are in four labeled small communities. We detected six
overlapping communities by starting from different ego components. Four of the detected communities are related to the four ground truth, as illustrated in (a) to (d).
 The green are the intersection of the detected community and the corresponding ground truth,
 while the pink and blue are the remainders of the ground truth and additional detected vertices.}
  \label{fig:multiMemberships_DBLP}
\end{figure*}

\subsection{All Local Membership Identification}

For the second and fundamental task of revealing all communities to which a single seed $s$ belongs,
we group the vertices in the ground truth according to their overlapping memberships $om$,
randomly pick 500 vertices in each group (pick all if the vertices in a group is less than 500)
and find all their membership communities.

 We preprocess the ground truth to remove identical copies of ground
truth communities, which are not relevant to questions of multiple memberships.
 After removing the identical communities,
 we have 1517, 4959, 4703, 4771 and 4885 ground truth communities
 for Amazon, DBLP, LiveJ, YouTube and Orkut.
 The ground truth communities of Amazon build a typical hierarchical dendrogram,
those of LiveJ also reveals a kind of hierarchical structure.
Others form overlapping relationships.

\begin{table}[t]
\begin{center}
\scalebox{1}{
\begin{tabular}{ l | l c c c c c | c } \hline
\bf{{Datasets}} &  $om$  &     1 &	   2   &   3  &   4	 &    5	  & Avg.  \\ \hline
    	  & Truth& 	0.850&	0.756&  0.661& 	0.583& 	0.439& 	0.659 \\
\bf{{Amazon}}&	Cond&	0.811& 	0.720& 	0.580& 	0.501& 	0.408& 	0.604 \\
          &  TPN&	0.824& 	0.746& 	0.600& 	0.506& 	0.382& 	0.616 \\
\hline							
    	  & Truth&  0.753	&0.650	&0.612	&0.558	&0.513	&0.6172\\
\bf{{DBLP}}&Cond&0.740	&0.624	&0.579	&0.509	&0.448	&0.580\\
          &  TPN&0.694	&0.616	&0.592	&0.489	&0.434	&0.565 \\
\hline							
    	  & Truth &0.671	&0.530	&0.435	&0.287	&0.180	&0.421 \\
\bf{{LiveJ}}&Cond & 0.601	&0.445	&0.356	&0.225	&0.166	&0.359 \\
           &  TPN&0.591	&0.45	&0.325	&0.229	&0.127	&0.345\\
\hline							
    	  & Truth&  0.411	&0.377	&0.313	&0.226	&0.163	&0.298  \\
\bf{{Youtube}} & Cond&0.429	&0.340	&0.254	&0.136	&0.132	&0.258\\
          &   TPN&0.389	&0.345	&0.236	&0.151	&0.120	&0.248 \\
\hline							
    	  &  Truth&0.315	&0.251	&0.201	&0.142	&0.097	&0.201 \\
\bf{{Orkut}}  &	Cond&0.224	&0.166	&0.180	&0.135	&0.09	&0.160\\
          &	 TPN&0.214	&0.166	&0.176	&0.129	&0.090	&0.155\\
\hline
\bf{{All}} &  Avg.&0.568	&0.479	&0.407	&0.320	&0.253	&0.405 \\
\hline
\end{tabular}}
\end{center}
\caption{\bf{F1 scores for multiple membership identification.}} 
\label{Multiple_membership}
\end{table}

Table \ref{Multiple_membership} shows the average F1 scores of $om \in \{1,2,3,4,5\}$
for the five datasets across domains
by using ground truth size or the best two scoring functions.
The average F1 scores decline for larger $om$,
 coinciding with the intuition that multiple memberships make accurate detection harder.

On average, we achieve a high F1 score of 0.40 over all datasets
and different stop conditions for the community boundary.
Note that our method is different from the method
of sweeping the function curve to find multiple local minima\cite{JaewonICDM2012},
which actually only finds hierarchical communities.

Fig. \ref{fig:multiMemberships_DBLP} illustrates all the detected local communities of a vertex, which is in four labeled ground truth communities, picked from the DBLP network.
We found six small overlapping communities and four of them are related to the ground truth communities.

\section{Conclusion}

We defined the local spectral basis of
a subspace spanned
by a few leading ``approximate eigenvectors" of the neighborhood subgraph around the seeds,
and present a systematic and effective approach called \emph{LOSP} for finding
local overlapping communities from subspaces.
Two observations are made regarding
the key parameters of \emph{LOSP}:
 the dimension of the local invariant subspace corresponds to the number of local structures,
and the steps of the random walks are related to the conductance and size of the target community.

For the semi-supervised learning task, finding a target community via very few exemplary seed members,
\emph{LOSP} outperforms the state-of-the-art local diffusion methods in real world networks across multiple domains,
and reduces the complexity by running in time and space
polynomial in the scale
of the local structure. 
We further effectively find all overlapping local communities for a single vertex by using \emph{LOSP} as a subroutine.

We define two new community scoring functions for the stoping rules of community boundary,
triad-participation-number (TPN) and normalized modularity,
and thoroughly evaluate them with three other existing metrics.
The five scoring functions behave well for \emph{LOSP},
and TPN actually performs better than conductance, which is regarded
as the current best by several arguments.

We also thoroughly investigate the structural properties of different seed sets,
and find that two types of seed structure better preserve the local structure.

\begin{itemize}
\item Low degree seeds are more ``loyal" to the local community in spreading out the random walk probabilities,
and better preserving
the random walk probability within the neighborhood of the seed set.
Random seeds behave similarly,
as most real networks follow a power law degree distribution.
\item Seeds that are highly cohesive to the target community, like high triangle participation seeds and low escape seeds,
also behave well.
\end{itemize}

A number of research issues remain to be addressed.
In future work,
we wish to answer how many seeds are required to have a unique
community of a given scale?
How can we
automatically determine the number of local structures
and hence the dimension of the local spectral subspace, and the steps of the short random walk?
We also wish to evaluate a combination of different seeding methods,
 like low degree plus high triangle participation.
We speculate that seeds with high diversity are of high quality, and we would like to explore
how to keep the diversity of the seeds such that each seed represents a unique portion of the community.
Another interesting work would be to find multiple local optimum
on the sorted probability vector to reveal local hierarchical structures of the small overlapping communities.

\section*{Acknowledgment}
This research work was supported by US Army Research Office W911NF-14-1-0477, and National Science Foundation of China 61472147.



%
%
%

{\scriptsize
\bibliographystyle{abbrv}
\bibliography{ICDMPaper.bbl}
}

\end{document}